\documentclass[11pt,a4paper]{article}
\usepackage{jinstpub}
\usepackage{lineno}
\usepackage{wrapfig}

\title{Design, production, burn-in and tests of the hybrid circuits of the Upstream Tracker at the LHCb detector}
\author[]{M.~Citterio,}
\author[]{N.~Conti,}
\author[1]{F.~De Benedetti,\note{\textcolor{black}{Corresponding author}}}
\author[]{P.~Gandini,}
\author[]{A.~Merli,}
\author[]{N.~Neri,}
\author[]{M.~Petruzzo}
\author[]{and E.~Spadaro Norella}
\affiliation[]{INFN Sezione di Milano,
Via Celoria 16, Milan, Italy}

\emailAdd{federico.debenedetti@mi.infn.it}

\abstract{We present a description of the design process, prototyping and production of the hybrid circuits for the front-end electronics of the Upstream Tracker at LHCb. The multilayer polyamide-based printed circuit boards, or hybrids,  are designed to host the front-end ASICs. The ASICs require an optimized power delivery network from 0 to 120MHz, with a maximum of $10^{-2}$ Ohms round-trip \textcolor{black}{resistance}, and 100 Ohms differential traces. Hybrids are required to have minimal radiation length, and to withstand the harsh environmental conditions of the data taking through intrinsic radiation hardness characteristics.}


\begin{document}
\maketitle

\section{Introduction}
The LHCb detector is undergoing a major upgrade phase after Run2  \cite{a}.
The upgraded LHCb detector will increase the luminosity of the previous runs by a factor of five.
All the trackers will be replaced by more performing detectors maintaining the general geometry of the experiment,
while  a new read-out \textcolor{black}{electronics systems} at 40MHz will be provided for all the sub-detectors \cite{b}.
The Upstream Tracker (UT) is the new silicon tracker in LHCb positioned before the magnet and will substitute the previous TT \textcolor{black}{(Tracker Turicensis)} detector.
It consists of four detection layers organized in staves that are covered with single-sided silicon strip sensors on both sides to avoid acceptance gaps. The two internal layers provide stereo information thanks to a stave rotation of ±5°. \textcolor{black}{The} UT employs four different sensor types to cope with the different occupancy, geometry, and radiation expected in the various \textcolor{black}{regions} of the detector. 
A dedicated 128 channels Application Specific Integrated Circuit (ASIC) called SALT (Silicon ASIC for LHCb Tracking) reads the incoming events from each sensor \cite{c}\textcolor{red}{\cite{h}}. A total of 4192 ASICs are mounted on flexible low mass polyamide circuits, called hybrids, providing power and digital signal routing. 
The first chapter shows the hybrid circuit design while the second one discusses their production.

\section{Hybrid circuit design}
\begin{figure}[htbp]
{
\centering
\includegraphics[width=1\textwidth]{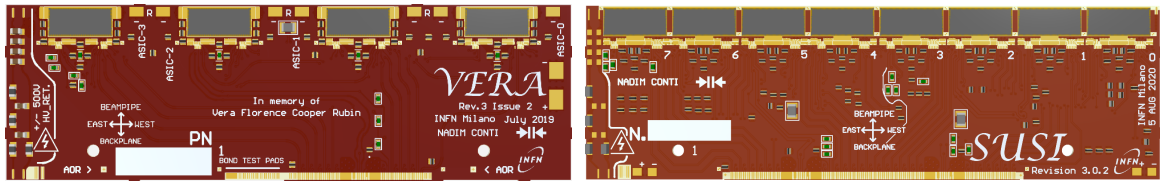}
}
\caption{\label{fig:SUSI-AND_VERA} On the right the 8-ASIC SUSI hybrid, on the left the 4-ASIC VERA version.}
\end{figure}
\begin{figure}[htbp]
{
\centering
\includegraphics[width=0.99\textwidth]{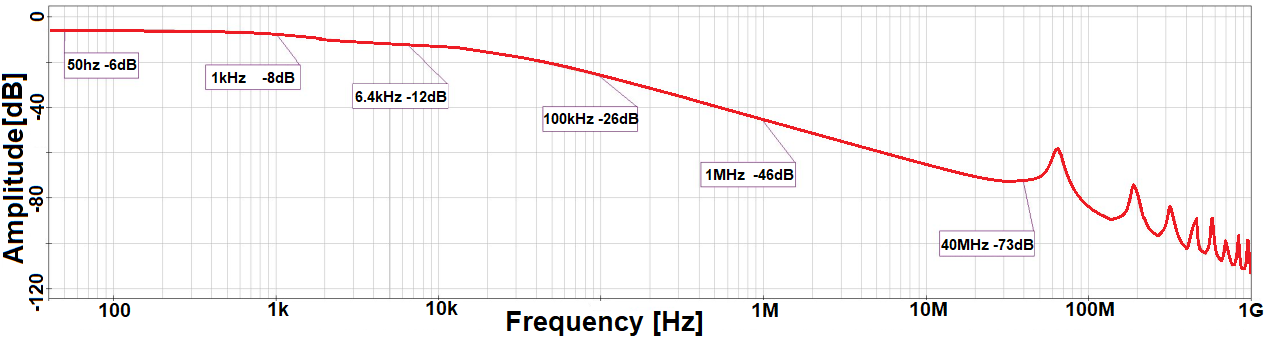}
}
\caption{\label{fig:HV_VERA_FILTER} Amplitude response of the high voltage network from the high voltage power supply to the sensor bonding pads.The attenuation is highlighted for significant frequencies, such as 40MHz.}
\end{figure}
\begin{figure}[htbp]
\centering
\includegraphics[width=.45\textwidth]{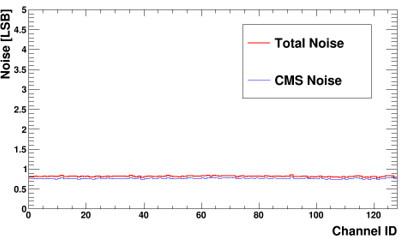}
\qquad
\includegraphics[width=.45\textwidth]{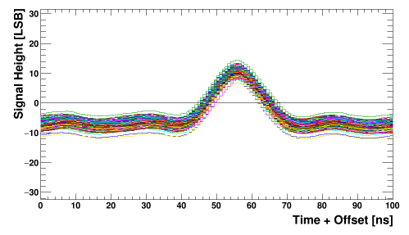}
\caption{\label{fig:prototest} \textcolor{black}{Electrical} tests on hybrid prototype. The figure on the left shows the noise plot, while the figure on the right shows the pulse scan plot.}
\end{figure}

The hybrid circuits, shown in figure \ref{fig:SUSI-AND_VERA}, are divided into two different versions. A 4-ASIC circuit (named VERA), designed for the outermost area of the detector, and an 8-ASIC version (named SUSI) designed for the innermost sector of the detector where the occupancy is the highest. The hybrids are polyamide-based, multilayer printed circuit boards with four copper layers and a final thickness of 320um for the 4 ASIC and 440um for the 8 ASIC version.
\begin{wrapfigure}{r}{0.55\textwidth}
{
\includegraphics[width=0.55\textwidth]{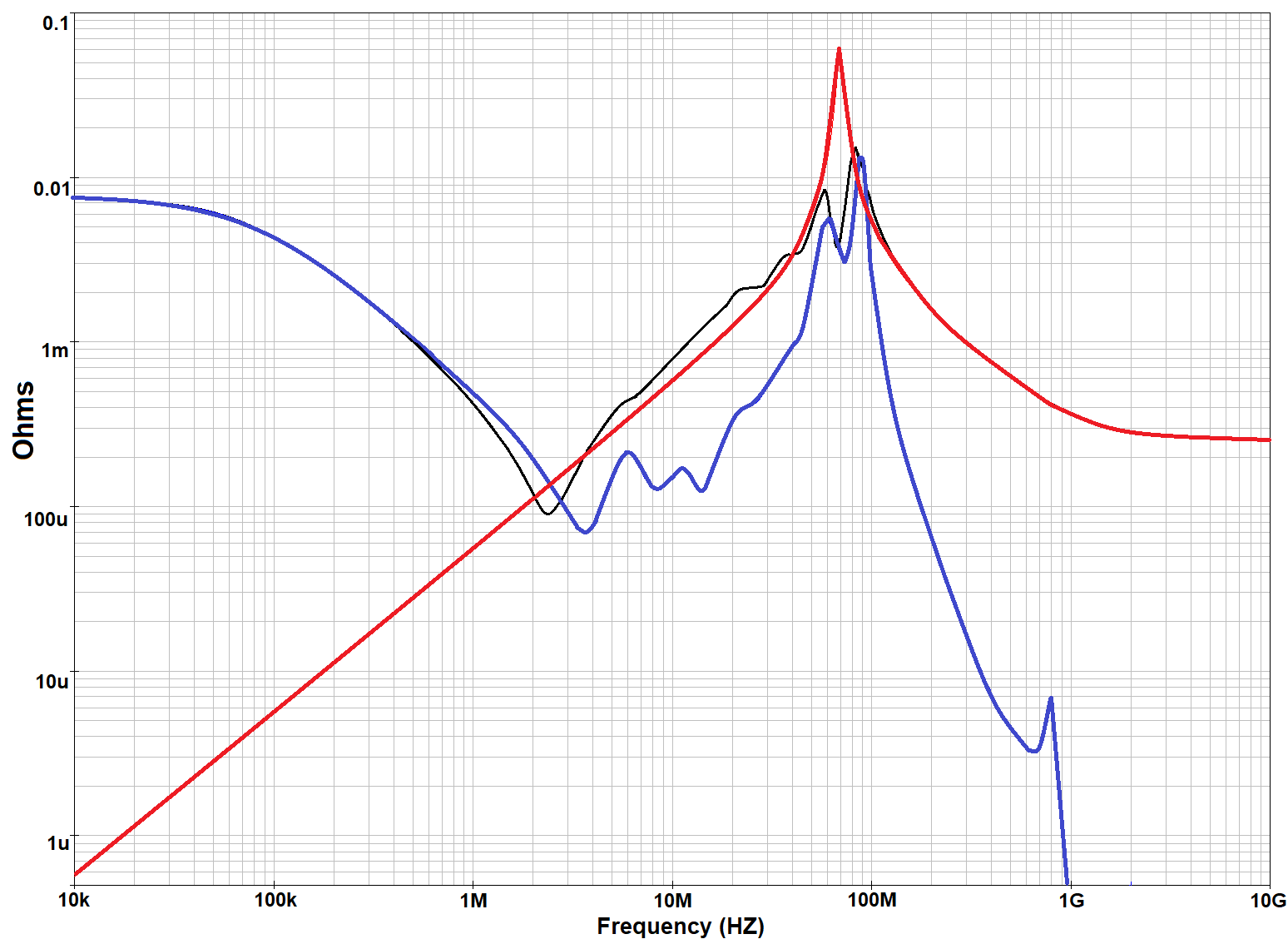}
}
\caption{\label{fig:PDN} The PDN response of the SUSI Hybrid is here detailed. In red, the ASIC on-die capacitance and bonding wires inductance resonance. In blue is the analogue domain behaviour and in black is the digital one.}
\end{wrapfigure}
The design is optimized to deliver 500mW of power per ASIC via the low voltage power lines and it is paired with a 500VDC sensor biasing network. A dynamic and static Power Delivery Network (PDN) analysis has been carried out to ensure proper power integrity for the low-voltage lines.
This analysis aims to avoid \textcolor{black}{excessive voltage} drop areas and to minimize the PDN's inductive behavior, features that could lead to a loss of reliability in the ASIC \cite{d}. A maximum peak below 20mOhms achieved in the dynamic PDN \textcolor{black}{impedance} helped to minimize baseline oscillations in the data stream and ensured that the ASICs would perform without glitches. To reach this goal, the usage of high-density wire \textcolor{black}{bonding} on the power pads, thin stack up dielectrics between the power planes and careful optimization of the \textcolor{black}{capacitor} values and their position on the PCB was mandatory \cite{e}. The high voltage biasing network is optimized to deliver a low pass filtering action to the sensor with a final -20dB attenuation at 10kHz and a maximum attenuation point before the first resonance at around 40 MHz (ASIC's clock frequency), as seen in figure  \ref{fig:HV_VERA_FILTER}. The filtering is also optimized to keep 6 nF of bulk capacitance next to the bonding wires biasing the sensor while maintaining a high safety factor regarding the dielectric isolation between high and low voltage lines (3x). Both hybrid versions were tested to the same 3x safety factor, equivalent to 1500 VDC, for the dielectric strength between the positive and negative high voltage lines.
A 4-Layer stack up with 18um thick metal layers is used to keep the radiation length low enough to meet the detector requirements \cite{a}. The final radiation length requirement is reached by minimizing the number of capacitors, the exposed gold-plated pads and by reducing the thickness of the plane separation dielectrics, which in turn improve, even more, the PDN inductance and peak impedance.
The VERA and SUSI hybrid prototypes have been manufactured and tested to assess their electrical \textcolor{black}{performance}. Both hybrid versions exceed the \textcolor{black}{detector} noise requirements, and \textcolor{black} {achieve} a noise level lower than 1 LSB, as shown in figure \ref{fig:prototest}. The pulse scan plot shows a residual digital pickup at 40 MHz.

\section{Hybrid circuit production}
\begin{figure}[htbp]
\centering
\includegraphics[width=.45\textwidth]{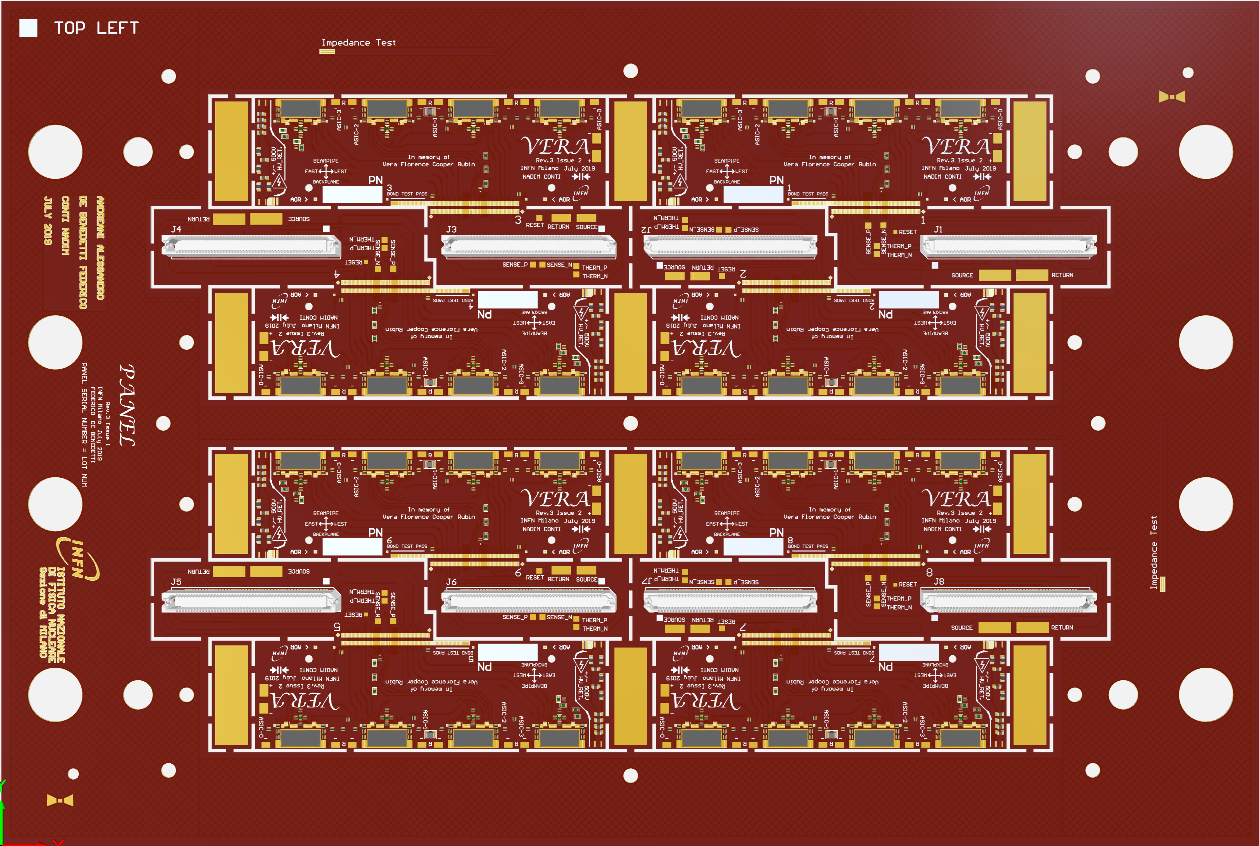}
\qquad
\includegraphics[width=.45\textwidth]{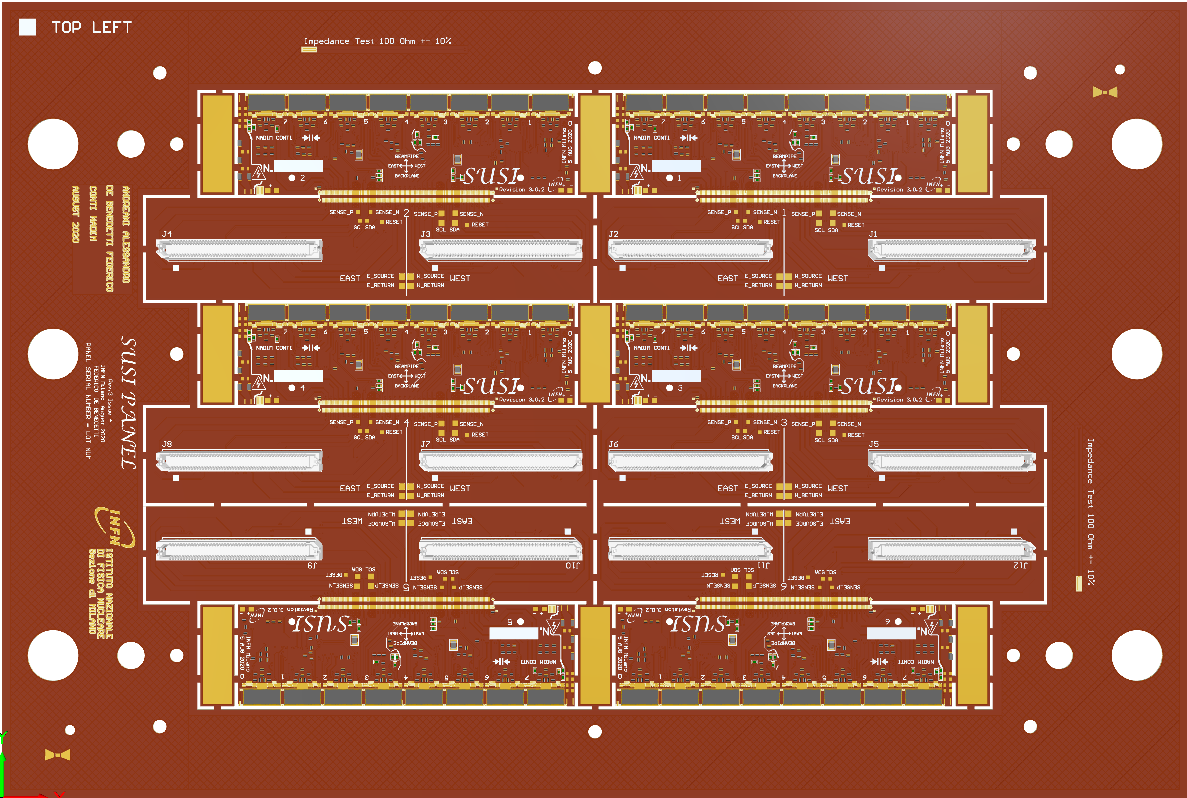}
\caption{\label{fig:hybrid panel design} Hybrid panels design. The right image shows the SUSI panel design with six single hybrids. The left image shows the VERA panel design with eight hybrids.}
\end{figure}

\begin{figure}[htbp]
\centering
\includegraphics[width=.55\textwidth]{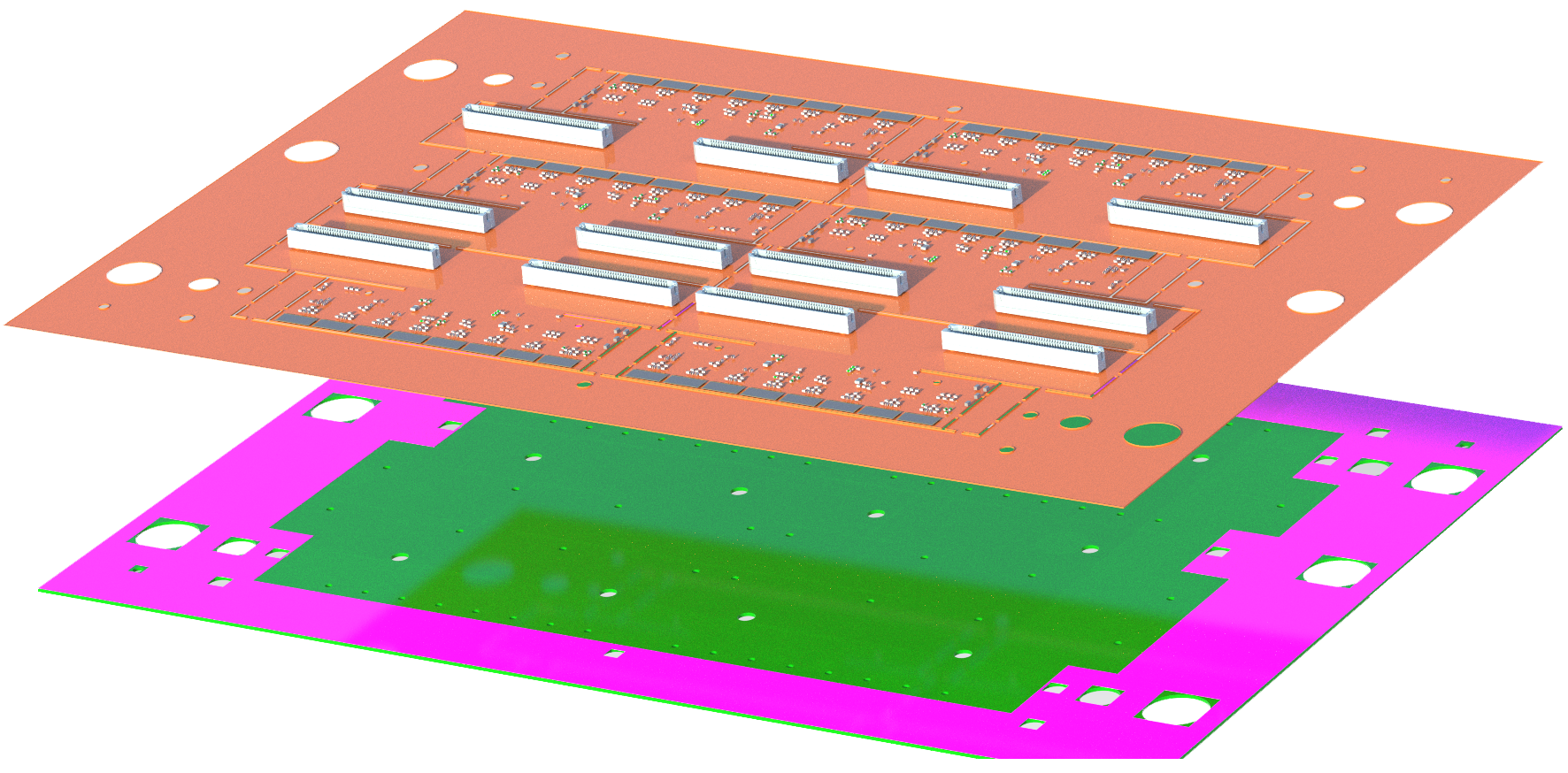}
\qquad
\includegraphics[width=.35\textwidth]{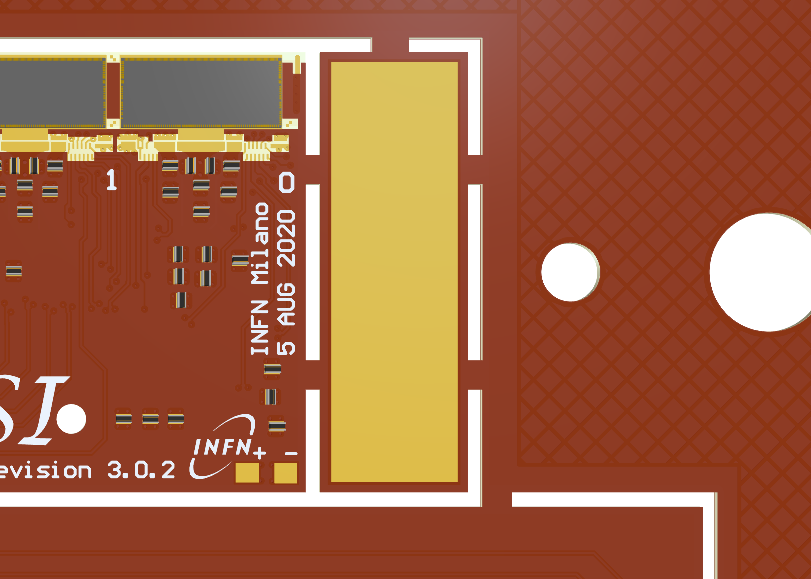}
\caption{\label{fig:FR4stackup} The left image shows the flex panel in brown, the FR4 stiffener in green and the glue pattern in pink. The right image shows  the \textcolor{black}{bridges} and voids around a single hybrid. }
\end{figure}

\begin{figure}[htbp]
\centering
\includegraphics[width=.29\textwidth]{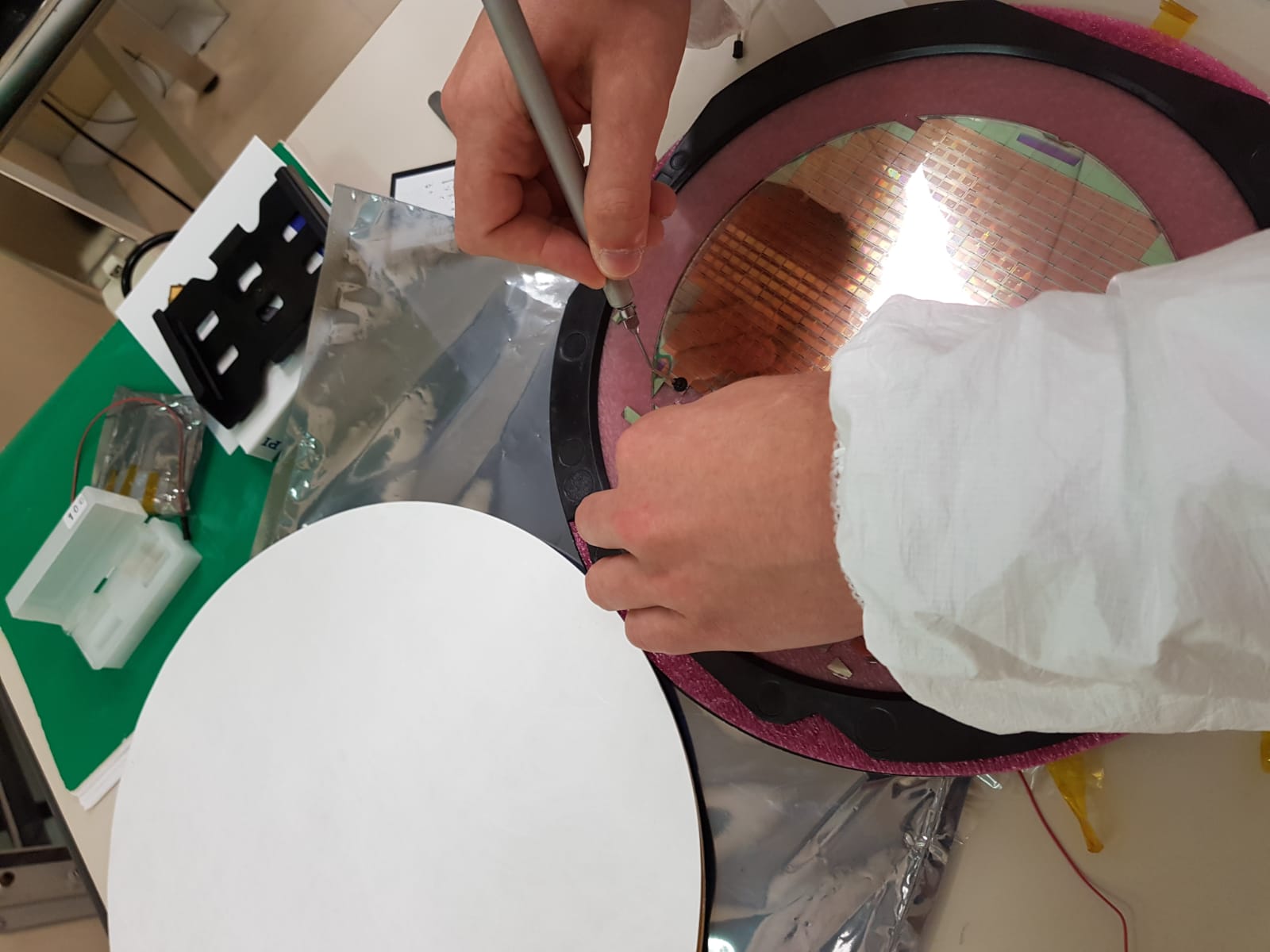}
\qquad
\includegraphics[width=.29\textwidth]{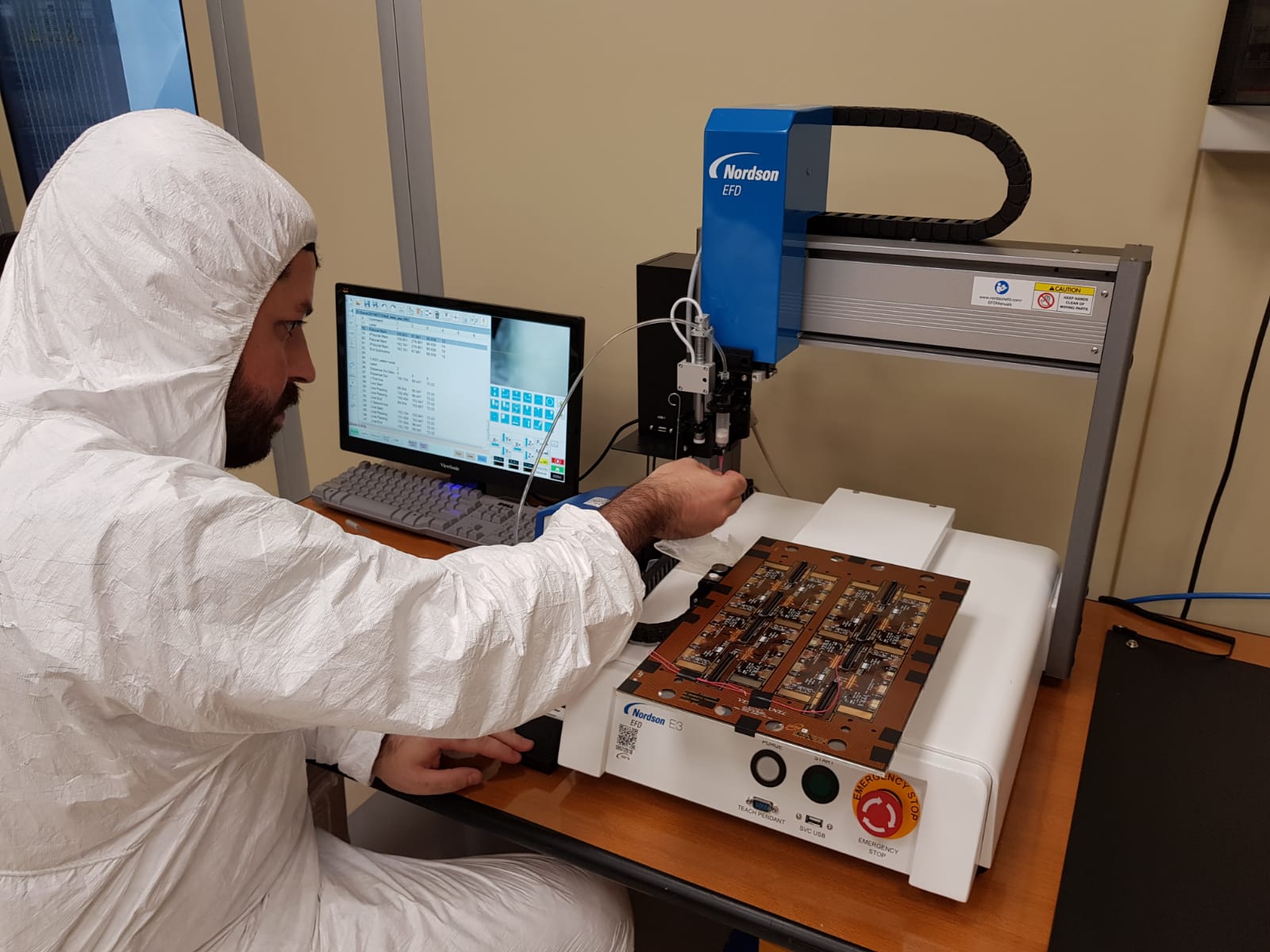}
\qquad
\includegraphics[width=.29\textwidth]{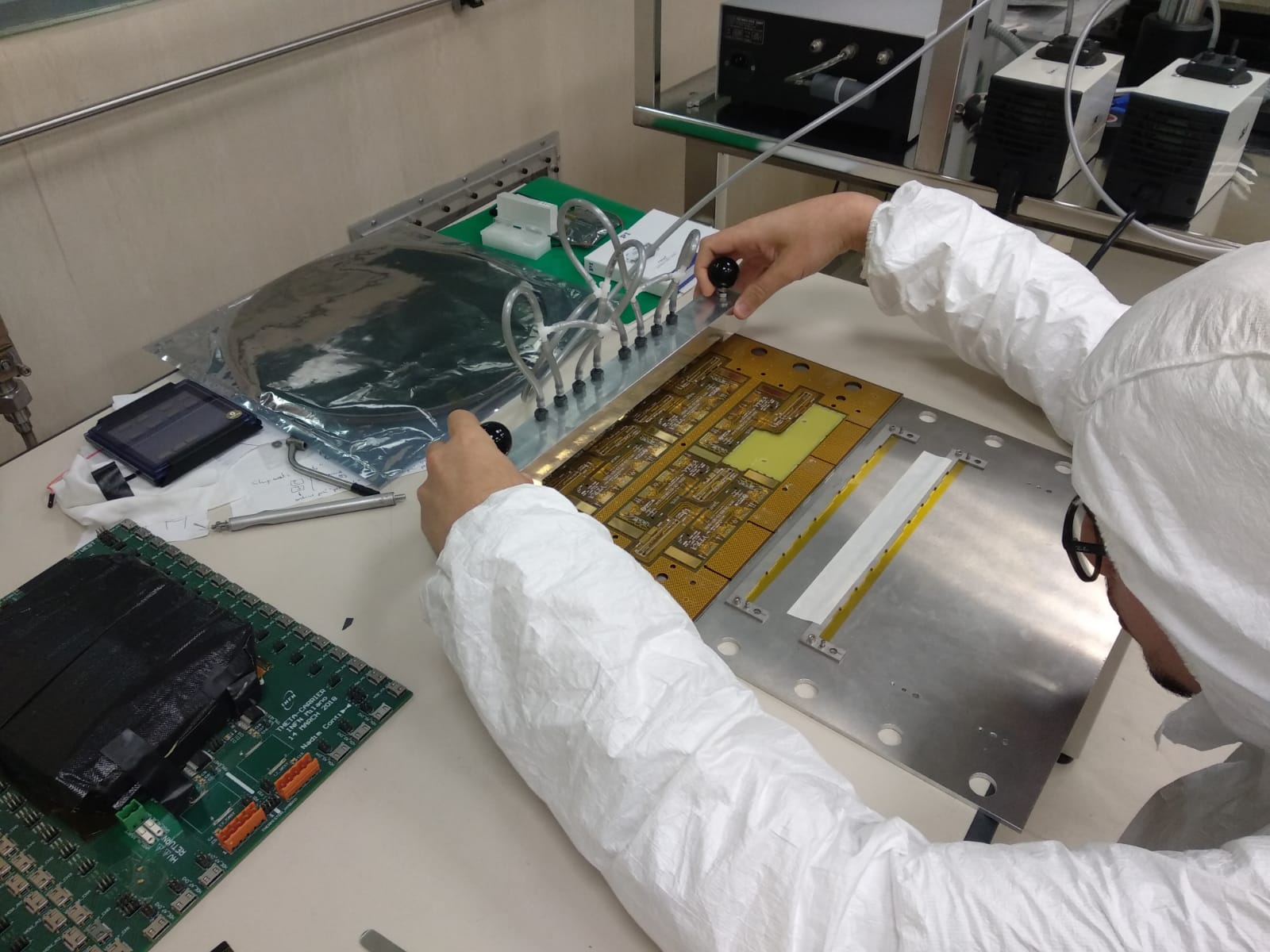}
\caption{\label{fig:production}The left figure shows the chip extraction operation from the wafer. The central figure shows the gluing operation with the robot. The right figure shows the ASICs alignment on VERA panel.}
\end{figure}

The hybrid circuit production involves several assembling and testing steps. Single hybrids are embedded into a panel to reduce production costs and assembly time. Two different designs of hybrid panels, one per hybrid version, have been manufactured, as shown in figure \ref{fig:hybrid panel design}.
Their size of 350x235mm allows embedding several copies of a single hybrid plus a dedicated connector for testing purposes.
The panels inherit the same stack up and material identified in the hybrid design.
A dedicated FR4 stiffener, \textcolor{black}{temporarily used during the assembly, is} glued to the back of the flex panel  \textcolor{black}{to improve} its mechanical strength.
The glue covers only the outermost region of the panel, as shown in figure \ref{fig:FR4stackup} (left), to allow cutting and removing operations during the module construction.
Many voids and bridges are added on the edge of a single circuit to improve the hybrid removal, as shown in figure \ref{fig:FR4stackup} (right). Several hole patterns allow aligning the different tools used during production and further processing. \textcolor{black}{The flex production and the passive component assembly
are carried out by an external company.}
The SALT chips, manufactured by TSMC, are organized in wafers. The electrical \textcolor{black}{performance} \textcolor{black}{is} assessed through a wafer test, which generates a list of good and bad ASICs and their positions. Only ASICs marked as good are extracted from the wafers and used, as shown in figure \ref{fig:production} (left). 
Chips are then glued with a bi-component silver-loaded resin, the Loctite Ablestik 2902, which has been chosen for its thermal and electrical conductivity. A dedicated robot guarantees to lay down the glue in the correct positions, following a precise glue pattern that allows to maximize the coverage and reduce glue spills. The glue has been tested for radiation hardness. 

The ASICs are finally aligned and mounted onto the hybrid panels. A custom vacuum pick-up tool is employed for VERA panels assembly with a precision of 100um.
An alignment surface hosts the ASICs that are then transferred to the hybrid panel using the vacuum pick-up tool, as shown in  figure \ref{fig:production} (right). Further corrections are performed manually under a microscope when needed.
For SUSI panels, the stringent alignment constraints require to execute the assembly with an automatic pick and place \textcolor{black}{machine}, \textcolor{black}{that positions the ASICs} to 10um of precision.
When the ASIC alignment is over, the curing takes place at ambient temperature for 24 hours.
The next production step consists of the wire bonding of the chips. Panels are bonded with a Delvotec 6400 G4 automatic machine. We use a 25um diameter (99\% Al + 1\% Si) wire, commonly used for this type of application.
A custom vacuum plate allows holding the panels during the operations. Pull tests performed during the production permit to assess the quality of the process.

After wire bonding, a burn-in test is performed. Hybrid panels are powered and driven in a climatic chamber, \textcolor{black}{where the maximum temperature is set to 60°C to prevent any damage to the capacitors mounted next to the ASICs, which can withstand a maximum temperature of 80°C. The burn-in test time is set accordingly to seven days to be effective \cite{g}}. 
A custom hardware was designed and \textcolor{black}{manufactured} to provide the nominal voltage required, a 40 MHz clock, and the register configurations via I2C to each hybrid. 
The \textcolor{black}{electronic} system can measure the temperature and the humidity of the environment. 
A dedicated online software checks the burn-in status and logs the events for offline analysis.
A final electrical test is performed on the single \textcolor{black}{hybrid} thanks to a dedicated electronic system provided by CERN, composed of a VLDB \textcolor{black}{\cite{f}} board and an acquisition card, called PCie40. The hybrid's \textcolor{black}{performance} \textcolor{black}{is} assessed by measuring several ASIC parameters, generating the same plots shown in figure \ref{fig:prototest}. A script analyzes the data and creates a detailed report, used for hybrid grading. Hybrids are grouped in four categories: A \textcolor{black}{(<2 bad channels)}, B \textcolor{black}{(<5 bad channels)}, C \textcolor{black}{(more than 5 bad channels)}, and F \textcolor{black}{(hybrid failure)}. Only A-graded hybrids are considered suitable for stave construction.
Hybrids with no problematic channels are the first to be selected for stave construction.

As a final step, each panel undergoes an optical inspection to detect mechanical problems. Typical issues are related to wire bonding, ASIC \textcolor{black}{damage}, or wire bond pad defects not spotted previously. 
The panels are then packed in 3D-printed custom made PLA transport boxes and shipped to the final assembly facility.
Each pack is sealed in a sealed bag to protect the electronics from moisture.
\section{Conclusion}
Two versions of polyamide-based hybrids for the Upstream Tracker have been designed and tested, \textcolor{black}{corresponding to different sensor types}.
A total number of 1080 (4-chip) VERA and 110 (8-chip) SUSI hybrids have been successfully manufactured, assembled, tested, and delivered for module and stave assembly. These numbers include the production of spare hybrids for future repairs.
The production of hybrids has been completed.
A first stave test in the final configuration is foreseen by the end of November 2021 at CERN, where 25 staves out of 68 have been already delivered. 
After assembling and testing the UT in the cleanroom, the aim is to install the detector underground by the end of February 2022.

\section{Acknowledgment}
We acknowledge support from the ERC Consolidator Grant SELDOM G.A. 771642

\end{document}